\documentclass[review, 11pt]{elsarticle}

\usepackage{lineno,hyperref}
\usepackage[latin1]{inputenc}
\usepackage[T1]{fontenc}

\usepackage{amssymb}          
\usepackage[amssymb]{SIunits}
\usepackage{amsmath}         
\usepackage{icomma}    
\usepackage{nicefrac}  
\usepackage{mathrsfs}  
\newcommand{\Tau}{\mathrm{T}}

\usepackage{a4wide,ae}
\usepackage{geometry}
\usepackage{graphicx}
\usepackage{subfigure}
\usepackage[margin=10pt,font=small,labelfont=bf]{caption}
\usepackage{float}

\usepackage{tabularx}
\usepackage{booktabs}
\usepackage{longtable}
\usepackage{multirow}

\journal{Astroparticle Physics}

\begin{document}

\begin{frontmatter}

\title{Improved $\gamma$/hadron separation for the detection of faint $\gamma$-ray sources using boosted decision trees}

\author[DESYaddress]{Maria Krause}
\ead{maria.krause@desy.de}

\author[UCDaddress]{Elisa Pueschel}
\ead{elisa.pueschel@ucd.ie}

\author[DESYaddress]{Gernot Maier}
\ead{gernot.maier@desy.de}

\address[DESYaddress]{DESY, Platanenallee 6, 15738 Zeuthen, Germany}
\address[UCDaddress]{School of Physics, University College Dublin, Belfield, Dublin 4, Ireland}

\begin{abstract}
Imaging atmospheric Cherenkov telescopes record an enormous number of cosmic-ray background events. Suppressing these background events while retaining $\gamma$-rays is key to achieving good sensitivity to faint $\gamma$-ray sources. The differentiation between signal and background events can be accomplished using machine learning algorithms, which are already used in various fields of physics. Multivariate analyses combine several variables into a single variable that indicates the degree to which an event is $\gamma$-ray-like or cosmic-ray-like. In this paper we will focus on the use of boosted decision trees for $\gamma$/hadron separation. We apply the method to data from the Very Energetic Radiation Imaging Telescope Array System (VERITAS), and demonstrate an improved sensitivity compared to the VERITAS standard analysis.
\end{abstract}


\end{frontmatter}


\section{Introduction}
\label{sec:intro}
Ground-based imaging atmospheric Cherenkov telescopes (IACTs) are used for the study of astrophysical objects emitting very-high-energy (VHE, E>\unit{100}{GeV}) $\gamma$-radiation. When a high-energy photon penetrates Earth's atmosphere, an electromagnetic cascade of secondary particles is generated. This cascade is called an extensive air shower. While traveling through the Earth's atmosphere towards the ground, the highly relativistic charged particles stimulate the emission of Cherenkov radiation. This emission is measured by IACTs. Multi-telescope arrays such as H.E.S.S. \cite{Aharonian:2006pe}, MAGIC \cite{Aleksic:2014aa, Aleksic:2014ab}, and VERITAS \cite{Holder:2008ux} provide a multi-dimensional view of the atmospheric showers. 

Cosmic-ray events also produce extensive air showers in Earth's atmosphere. This leads to a high background rate for IACTs. Under normal operating conditions, VERITAS triggers at a rate of $\sim$350 Hz, while the number of $\gamma$-rays recorded even for a strong $\gamma$-ray emitter is less than 1 per second. In order to achieve sensitivity to weak $\gamma$-ray sources, it is essential to suppress the background events while retaining the $\gamma$-rays associated with astrophysical objects. In the energy ranges of the currently operating IACTs, the cosmic-ray spectrum is dominated by protons. Thus, the shower properties of hadronic cosmic rays are considered in the following, and the background due to electrons is not specifically addressed. 

The shower images from $\gamma$-rays and hadronic cosmic rays have slightly different properties. Cosmic-ray images have more irregular shapes than $\gamma$-ray images, due to the development of the cascade via hadronic production of particles with large transverse momentum. In addition, substructures due to the creation of electromagnetic subshowers in hadronic air showers are visible. By contrast, the electromagnetic cascades induced by $\gamma$-rays result in an even distribution of energy between the secondary particles, leading to compact, regular images. 

The showers can be described by a series of properties that differ, on average, between the signal $\gamma$-ray events and the background cosmic-ray events, referred to in the following as \textit{training~parameters}. Events can be statistically separated into signal and background events by making selection requirements on the training parameters. This can be done for the training parameters individually (\textit{standard~box~cuts}), or with multivariate analysis techniques such as Boosted Decision Trees (BDTs) \citep{Breiman:1984crt, Freund:1999bt}. Multivariate techniques combine several training parameters into a single discriminating variable indicating the degree to which an event is $\gamma$-ray- or cosmic-ray-like. BDTs take into account nonlinear correlations between training parameters and ignore weak training parameters, giving the technique advantages in power and robustness over several other multivariate methods~\cite{Mitchell:1997}. 

In the present study, it will be shown that the BDT method provided by the Toolkit for Multivariate Data Analysis (TMVA) package \cite{TMVAGuide} can be used to improve the discrimination between $\gamma$-ray and cosmic-ray events for IACTs. The performance and the stability of the method will be assessed through application to data from $\gamma$-ray sources observed by VERITAS, an array of four IACTs located at the Fred Lawrence Whipple Observatory in Southern Arizona~\cite{Holder:2008ux, Weekes:2001pd}. The BDT technique has previously been demonstrated to yield sensitivity improvements in analysis of data from the H.E.S.S. array~\cite{Ohm2009, Becherini:2011pb}. A closely related technique, the random forest method, is used in the analysis of data from the MAGIC array~\cite{MAGICrandomforest}.

\section{Classification using boosted decision trees}\label{sec:algorithm}
BDTs are based on a simpler object, the decision tree. In order to construct a decision tree, a training sample and a set of training parameters are used. The training sample is a mixture of signal and background events, where every event's type is known (denoted as $Y_{i}=+1$ for signal, $Y_{i}=-1$ for background). The set of training parameters is used to discriminate between signal and background. A tree is built by making a series of binary splits of the training sample into nodes of increasing signal and background purity. The division of events in the previous node is achieved by choosing both 1) the training parameter and 2) the value of the cut on this training parameter for which the separation between signal and background is maximized. The training of a tree is stopped when the number of events in a leaf is smaller than a predefined value, or the signal/background purity of a leaf exceeds a predefined value. The final nodes (\textit{leaves}) are designated as signal leaves if they are signal-dominated and background leaves if they are background-dominated.  

A disadvantage of decision trees trained in this way is the sensitivity to statistical fluctuations in the training sample. \textit{Boosting} is an iterative method to stabilize the performance \cite{Freund:1999bt, Schapire:1999ab, Roe:2004ya}. This process requires training a \textit{forest} of multiple decision trees. At the beginning of the training, all events have the same weight, $\omega_{i}$, and the tree, $t_{0}$, is built as described above. A misclassified event is an event classified as the wrong type, e.g.~a signal event assigned to a background leaf  or a background event assigned to a signal leaf. 

At each iteration of the training after the first tree, the weight of all misclassified events is increased by a boost factor, ${\alpha_t}$. In the following, the AdaBoost method \cite{Freund:1999bt, Schapire:1999ab} is used, where ${\alpha_t}$ is computed by

\begin{equation}
\alpha_{t}=\beta\cdot\ln\left(\frac{1-\epsilon_{t}}{\epsilon_{t}}\right).
\end{equation}
The parameter, $\beta$, is the user-specified learning rate and $\epsilon_{t}$ is the weighted fraction of misclassified events in the previous tree $t_{i-1}$. The weight of a misclassified event is thus given by

\begin{equation}
\omega_{i}=\omega'_{i}\cdot \exp{\alpha_{t}}
\end{equation}
where $\omega'_{i}$ is the weight of event $i$ in the previous tree. The re-weighting forces the training of the next tree to focus on events which were not classified correctly in the previous iteration. For the purposes of normalization, all events of the previous tree are re-weighted by

\begin{equation}
\omega_{i}=\frac{\omega'_{i}}{\sum\limits_{Y_{i}\neq T_{t}(x_{i})} (\omega'_{i}\cdot \exp{\alpha_{t}})+\sum\limits_{Y_{i}=T_{t}(x_{i})} \omega'_{i}}
\end{equation}
where $\Tau_{t}(x_{i})$ is either +1 if the event $x_i$ is classified as a signal event or --1 if the event is classified as a background event. A correctly classified event is represented by $Y_{i}=\Tau_{t}(x_{i})$ and a misclassified event by $Y_{i}\neq T_{t}(x_{i})$. A forest of $N_{Tree}$ trees is trained according to this process. The test sample is scored to determine the response of the BDT. This is performed by summing over all $N_{Tree}$ trees with

\begin{equation}
\Tau(x_{i})=\sum\limits_{t=1}^{N_{Tree}}\alpha_{t}\Tau_{t}(x_{i}),
\end{equation}
where $\Tau(x_{i})$ is the output or response variable of the BDTs. Using the information from the training stage, each event in a new dataset is assigned a value of $\Tau(x_{i})$. Then, this parameter is compared to the optimized cut value described in Section~\ref{sec:optimization}.

The BDT method used in this study is provided by the TMVA package, which is part of the ROOT data analysis framework (TMVA version 4.2.0, ROOT version 5.34.14 \citep{ROOT:1997}). The following BDT settings are used within this work:
\begin{itemize}
 \item The learning rate of the misclassified events in a tree ($\beta$ or \textit{AdaBoostBeta}) was set to 1 following the TMVA default value. This factor is used in the computation of the boost factors.
 
 \item The pruning method \textit{CostComplexity} was used \citep{Breiman:1984crt}. Pruning reduces statistical fluctuations by removing insignificant branches. In the studies presented here, the combination of pruning and deeper trees leads to a better separation than no pruning and shallow trees. Deeper trees have the advantage that all training variables are used. Pruning is necessary to stabilize performance when growing deeper trees.

 \item The separation type was chosen to be \textit{GiniIndex}. This parameter computes the inequality between signal and background distributions. 

 \item The number of required events for training and testing refers to the number of events remaining after the preselection cuts (described in Section~\ref{sec:sample}). The number of signal events is set equal to the number of background events.
\end{itemize}

The selection of the remaining BDT parameters (the number of trees, the minimum leaf size, and the maximum training depth) is described in Section~\ref{sec:BDTparameters}.

\section{Training of the boosted decision trees}

\subsection{Training parameters}

\begin{figure}
\centerline{\includegraphics[width=1.0\textwidth]{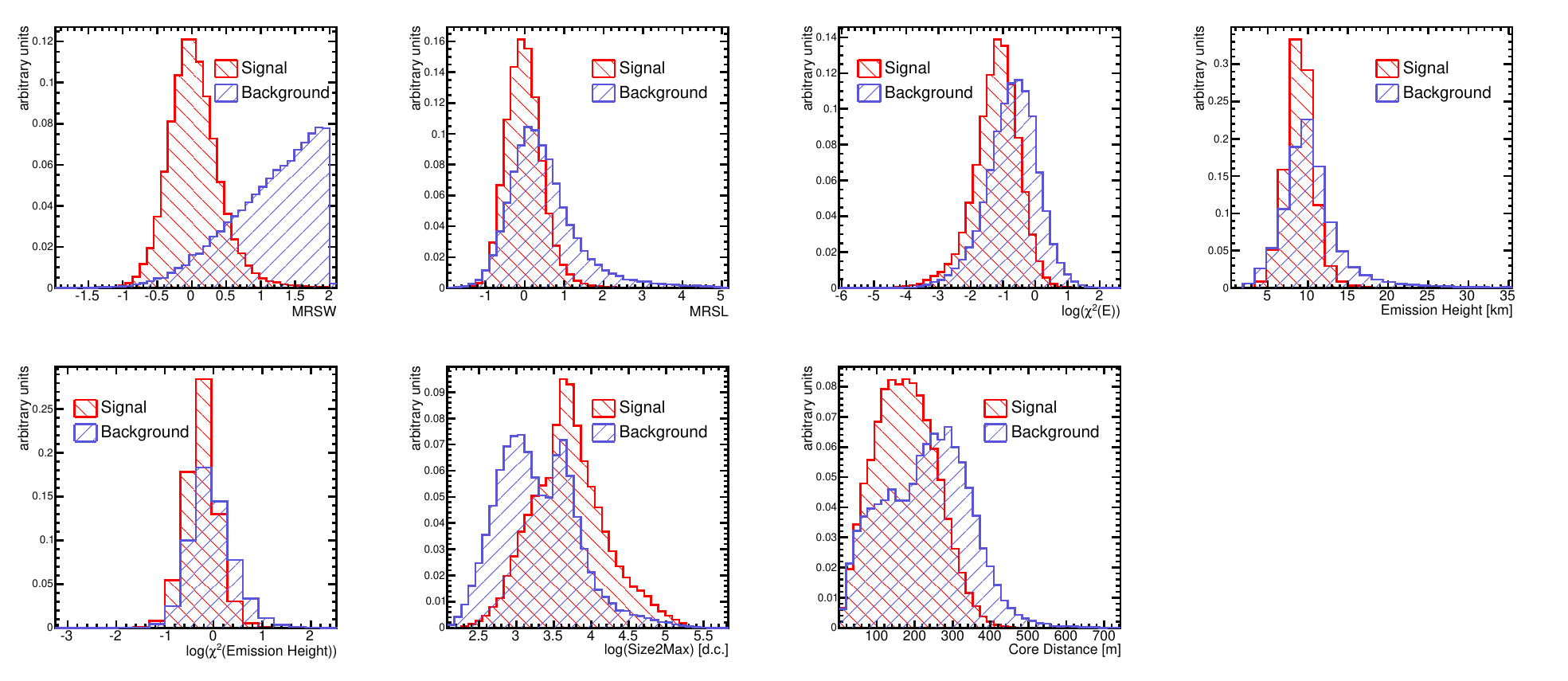}}
\caption{Parameters used in the BDT training, for the input signal and background training samples. The distributions shown are for events with energies greater than \unit{1}{TeV} and zenith angles of observation between 0$^{\circ}$ and 22.5$^{\circ}$.}\label{fig:TrainPars}
\end{figure}

The canonical method for parameterizing the images recorded by IACTs is to use Hillas parameters~\cite{Hillas, Fegan1997}, which describe the moments of the image ellipses. The measured width and length depend on the energy of the primary particle, the impact parameter $R$ and the level of the night-sky-background (NSB). Lookup tables generated from $\gamma$-ray Monte-Carlo simulations are used to relate the measured shower properties to the $\gamma$-ray energy. The image width and length for individual telescope images are combined in a weighted average for an array of telescopes with multiple images per event. The derived parameters are called the mean reduced scaled width (\textit{MRSW}) and the mean reduced scaled length (\textit{MRSL}). They are derived as

\begin{equation}
MRSW=\frac{1}{N_{images}}\sum_{i=0}^{N_{images}}\left(\frac{w_{i}-\hat{w}(R,s)}{\sigma_{w_{MC}(R,s)}}\right)\cdot\left(\frac{\hat{w}(R,s)}{\sigma_{w_{MC}(R,s)}}\right)^2
\end{equation}
and
\begin{equation}
MRSL=\frac{1}{N_{images}}\sum_{i=0}^{N_{images}}\left(\frac{l_{i}-\hat{l}(R,s)}{\sigma_{l_{MC}(R,s)}}\right)\cdot\left(\frac{\hat{l}(R,s)}{\sigma_{l_{MC}(R,s)}}\right)^2,
\end{equation}
where $s$ is the size of the image in digital counts, obtained by summing the charge of all the pixels of the image. The number of digital counts in the image depends on the pixel cleaning thresholds (described in e.g. \cite{Daniel2007}). The parameters $\hat{w}$ and $\hat{l}$ characterize the median, whereas $\sigma_{w_{MC}}$ and $\sigma_{l_{MC}}$ define the 90\% width values of the expected distribution of the image width and length, respectively. While the mean-scaled parameter distributions for $\gamma$-ray showers are centered around zero, hadronic showers tend to produce wider and longer images. In particular, the average \textit{MRSW} for hadronic showers increases dramatically as the shower energy increases, making this parameter a powerful discriminator. This is demonstrated in the first two plots of Figure \ref{fig:TrainPars}.

The size of the second largest image in digital counts, \textit{Size2Max}, is also used to discriminate between $\gamma$- and cosmic rays. This variable has a similar separation power to the size of the largest image, but is more stable against shower-to-shower fluctuations. Showers from $\gamma$-rays are on average brighter than cosmic-ray showers, resulting in a larger image size.

Several other quantities are useful for $\gamma$/hadron separation. Used here is the emission height, which is calculated for all pairs of telescopes and combined into a size-weighted average. Also used is the distance from the center of the array to the position of the shower core  on the ground.

The $\chi^{2}$-values for the energy and emission height have different distributions for $\gamma$- and cosmic rays. The $\chi^{2}$-values for the energy reconstruction are calculated based on scatter of the energy estimates in the individual telescopes from the size-weighted average energy estimate. The emission height  $\chi^{2}$ is similarly derived from the scatter of the pair-wise estimation of the emission height from the size-weighted average value. On average, cosmic-ray events exhibit larger energy and emission height $\chi^{2}$-values than $\gamma$-ray events.

\begin{figure}
\centerline{\includegraphics[width=1.0\textwidth]{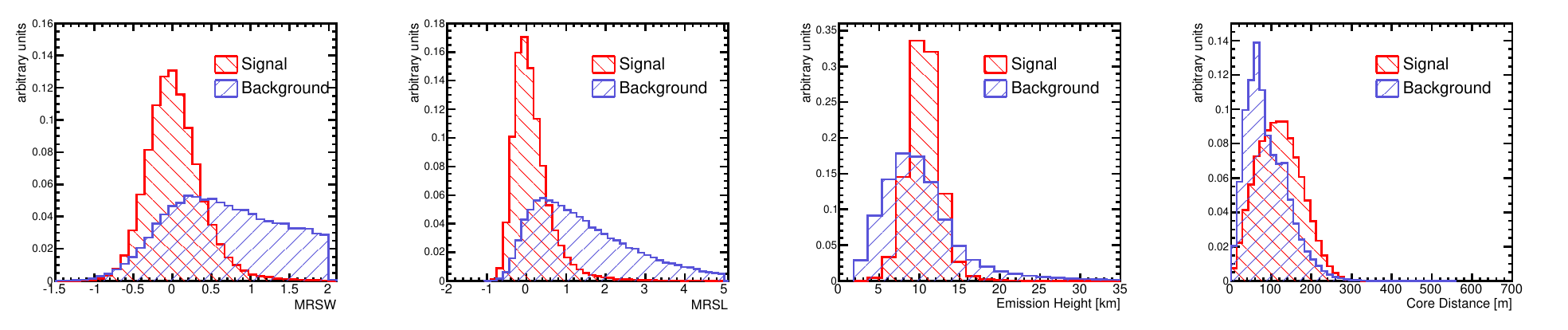}}
\centerline{\includegraphics[width=1.0\textwidth]{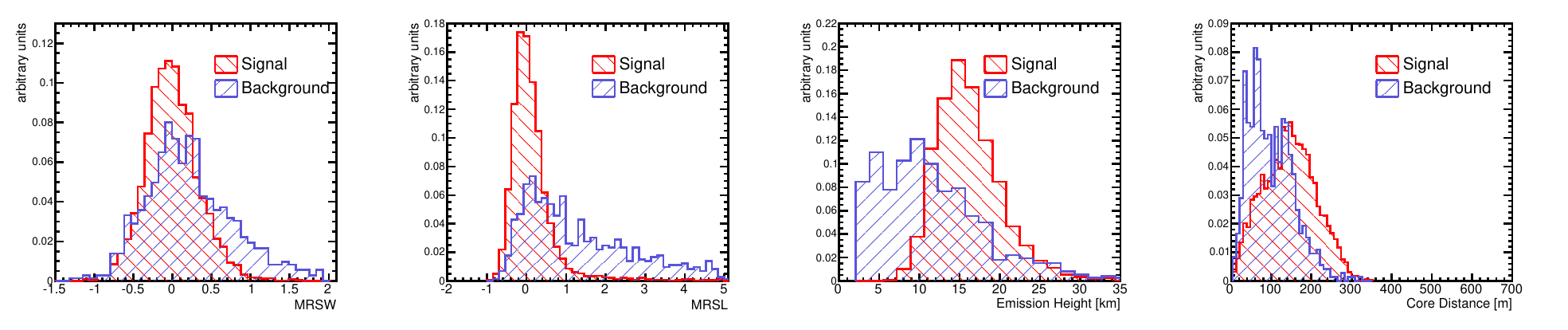}}
\centerline{\includegraphics[width=1.0\textwidth]{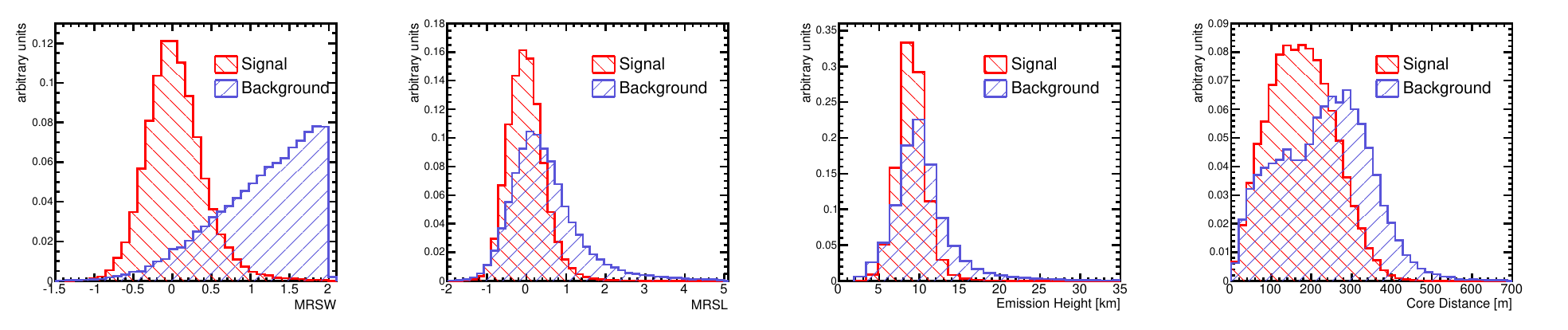}}
\caption{The four training parameter distributions that vary most with energy and zenith shown for different energy/zenith angle bins. \textit{Top} panel: energy between 0.08 and 0.32 TeV and zenith angle of observation between 0$^{\circ}$ and 22.5$^{\circ}$. \textit{Middle} panel: energy between 0.08 and 0.32 TeV and zenith angle of observation greater than 42.5$^{\circ}$. \textit{Bottom} panel: energy greater than \unit{1}{TeV} and zenith angle of observation between 0$^{\circ}$ and 22.5$^{\circ}$.}\label{fig:EnergyZenithDependence}
\end{figure}

Summarizing, the training parameters used are: \textit{MRSW}, \textit{MRSL}, log($\chi^{2}$(E)), emission height, log($\chi^{2}$(emission height)), log($Size2Max$), and the distance from the array center to to the shower core on the ground. Fig.~\ref{fig:TrainPars} shows example distributions of these parameters. The distributions shown are for events with energies greater than 1 TeV and zenith angles of observation between 0$^{\circ}$ and 22.5$^{\circ}$. The most powerful discriminating variable for this energy and zenith angle range is clearly the \textit{MRSW}. Several of the training parameter distributions vary with energy and zenith angle of observation as demonstrated in Fig.~\ref{fig:EnergyZenithDependence}. Accordingly, the BDT training is performed in bins of energy and zenith angle. Four energy bins (\unit{0.08--0.32}{TeV}, \unit{0.32--0.50}{TeV}, \unit{0.50--1.00}{TeV} and $>$\unit{1.00}{TeV}) and four zenith angle bins (0$^{\circ}$--22.5$^{\circ}$, 22.5$^{\circ}$--32.5$^{\circ}$, 32.5$^{\circ}$--42.5$^{\circ}$ and $>$42.5$^{\circ}$) are used, with bin widths selected to allow comparable event counts in each bin. 

\subsection{Signal and background estimation with VERITAS}
Observations with VERITAS are taken in the so-called \textit{wobble} mode~\cite{Fomin1994}, pointing the telescope at some offset (nominally 0.5$\degree$) away from the target. This allows simultaneous estimation of the emission from the source candidate (ON-region) and the background (OFF-region). The parameter $\alpha$ indicates the ratio between the acceptance of the ON- and OFF-regions. The number of excess (signal) events can be obtained from an ON-region and one or multiple OFF-regions using the reflected region or ring background method (see~\cite{Berge2007} for details). The excess events are computed by $N_{\gamma}=N_{ON}-\alpha\cdot N_{OFF}$, where $N_{ON}$ and $N_{OFF}$ are the number of events for the ON- and OFF-region, respectively.

\subsection{Choice of the training samples}\label{sec:sample}

The BDTs are trained with known signal and background samples: simulated $\gamma$-ray events for the signal, and OFF-source events from data for the background. These data were taken from observations of point sources with only one known $\gamma$-ray source candidate in the field-of-view of the camera. To avoid contamination of the background sample with $\gamma$-rays, only events with a shower direction of $>$0.22$^{\circ}$ from the expected source position in the center of the camera are used. The NSB levels observed in data vary substantially depending on the location of the source candidate in relation to the Galactic plane. For observations of regions with high NSB levels, pixels located close to the edge of the image may be removed during image cleaning as the cleaning threshold is based on the ratio of measured charge per pixel to expected variation due to NSB light. Thus, the NSB level impacts the \textit{MRSW} and \textit{Size2Max} distributions. Therefore, it is necessary to train the BDTs over the full range of NSB levels observed in data. A mix of galactic (high NSB level) and extragalactic (low NSB level) fields are used for the background training sample, and the NSB levels in simulation are selected to match the range of NSB levels found in the background training sample. Data used for the background training sample were collected under good weather conditions, with all four telescopes operational, and with a wobble offset of 0.5$^{\circ}$.

The VERITAS instrument has been upgraded twice. One of the telescopes was relocated during summer 2009 to make the array more symmetric~\cite{Perkins:2009T1}. In summer 2012, the photomultiplier tubes of each camera were upgraded~\cite{Otte:2011vu, Kieda:2013aa, Park:2015vp}. The changes in the VERITAS array impact the training parameter distributions, necessitating three separate BDT trainings that use signal and background samples from the appropriate time periods. 

Earth's geomagnetic field affects shower development, thus it is expected that the shower parameters will vary with the azimuthal angle of observation. However, separate trainings were not performed for observations of northern versus southern source candidates, as further separating the selected training samples resulted in an inadequate number of background training events for southern observations. Similarly, it is expected that the shower parameters will vary with the atmospheric conditions, but separating the selected training samples into winter and summer observations resulted in too few background training events for summer observations. The final training was performed without subdividing the training sample by season. However, the selected training sample can be extended in the future to include more summer and southern observations, enabling finer subdivisions.

Before the BDTs are trained, preselection requirements are made for \textit{MRSW} (--2.0 $<$ \textit{MRSW} $<$ 2.0) and \textit{MRSL} (--2.0 $<$ \textit{MRSL} $<$ 5.0). These requirements remove trivially classifiable background events, reducing the background sample to events that are difficult to distinguish from $\gamma$-rays. Images far from the camera center at a distance larger than 0.78$^{\circ}$ were also removed to avoid distortion effects.

\subsection{BDT training options}\label{sec:BDTparameters}

The shape and separation of the BDT response for signal and background depends on the specifications that the user sets on the individual trees and on the BDT forest. The BDT training options are selected with the goal of maximizing the separation between the signal and background response distributions, while avoiding overtraining. To test for overtraining, the BDT response is compared for the training sample and an independent test sample. The agreement between the test and training response distributions are quantified with a Kolmogorov--Smirnov test. A non-zero result of this test indicates that the BDTs are not overtrained.

The effect of varying the number of trees ($N_{Tree}$), the minimum number of events in a leaf (leaf size, \textit{MinEvents}), and the maximum training depth (\textit{MaxDepth}) on the BDT response was studied. The number of trees in the forest did not affect the separation between the signal and background response distributions for $N_{Tree}\geq$200, thus 200 trees were used in the training to reduce the computation time. The separation of the signal and background response distributions was found to increase with increasing \textit{MaxDepth} and with decreasing \textit{MinEvents}. However, too deep of a training and too small of a minimum leaf size resulted in overtraining. Values of \textit{MaxDepth} (\textit{MaxDepth}=50) and \textit{MinEvents} (\textit{MinEvents}=100) that avoided overtraining in all energy and zenith angle bins were selected.

\subsection{Training output and optimization}\label{sec:optimization}

\begin{figure}
\centerline{
\includegraphics[width=0.5\textwidth]{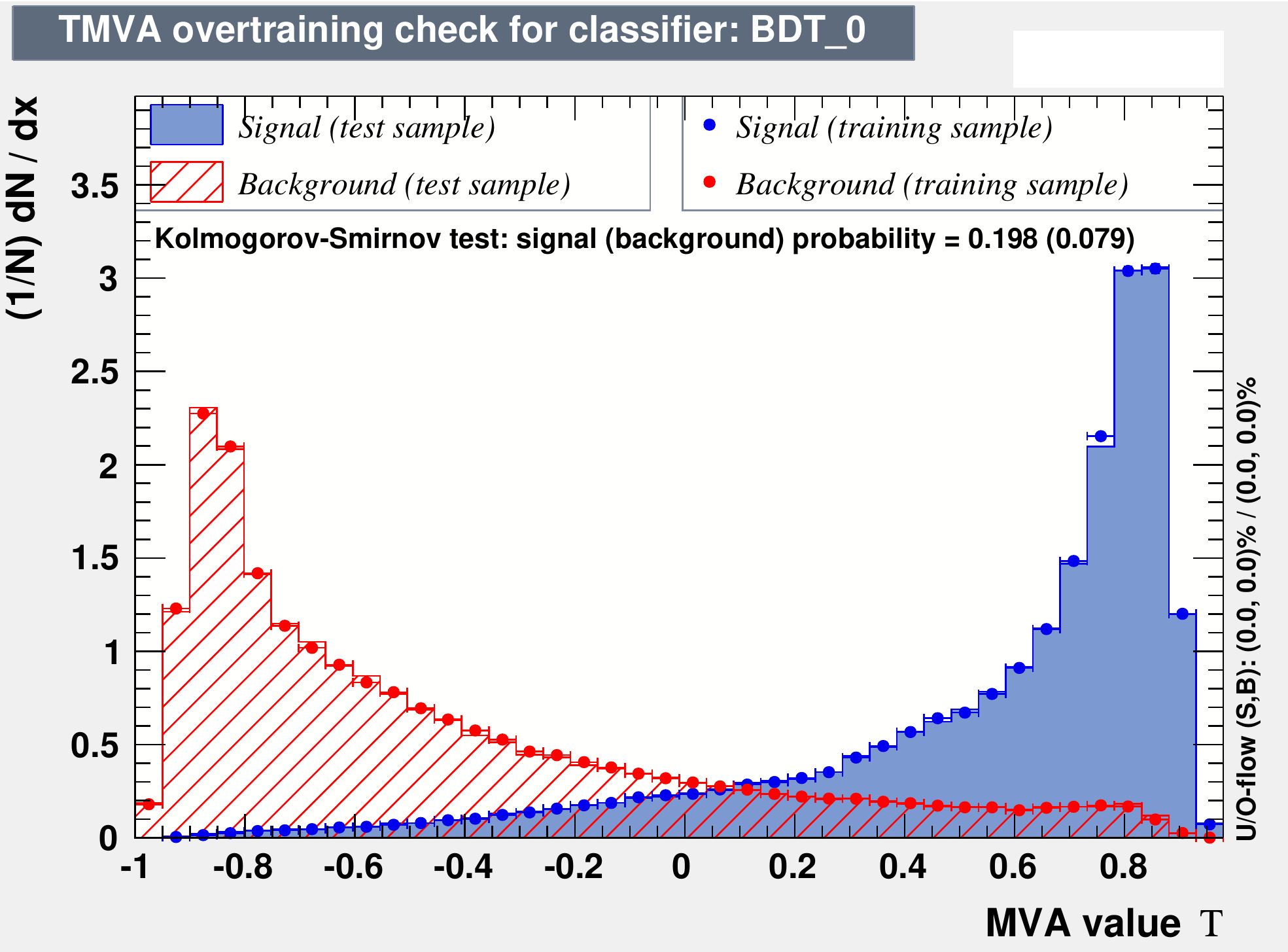}}
\caption{Comparison of training and test sample for both signal (blue, filled) and background (red, striped) events. Events in the sample have energies of \unit{0.08-0.32}{TeV} and zenith angles of 0$^{\circ}$--22.5$^{\circ}$.}\label{fig:BDT Overtraining}
\end{figure}

The BDT response $\Tau$ was compared for the training sample and a test sample in each energy and zenith angle bin. Good agreement was observed in all bins, indicating that the BDTs are not overtrained. As an example, this is shown in Fig.~\ref{fig:BDT Overtraining}, which displays $\Tau$ for the training sample and an independent test sample of known signal and background events of energy \unit{0.08-0.32}{TeV} and zenith angle 0$^{\circ}$--22.5$^{\circ}$. 

An optimal selection requirement $\Tau_{sel}$ on the BDT response is determined in each energy and zenith angle bin, such that events with $\Tau$ above (below) $\Tau_{sel}$ are considered $\gamma$-rays (cosmic rays). Three $\Tau_{sel}$ sets are determined for use on different types of objects. The types are defined by the flux and the spectral index $\Gamma$ of the object. The corresponding $\Tau_{sel}$ sets are for soft ($\Gamma\lesssim-3.5$), moderate ($\Gamma$ of -2.5 to -3.5), and hard source candidates ($\Gamma\gtrsim-2.5$).

\begin{figure}
\centerline{\includegraphics[width=0.5\textwidth]{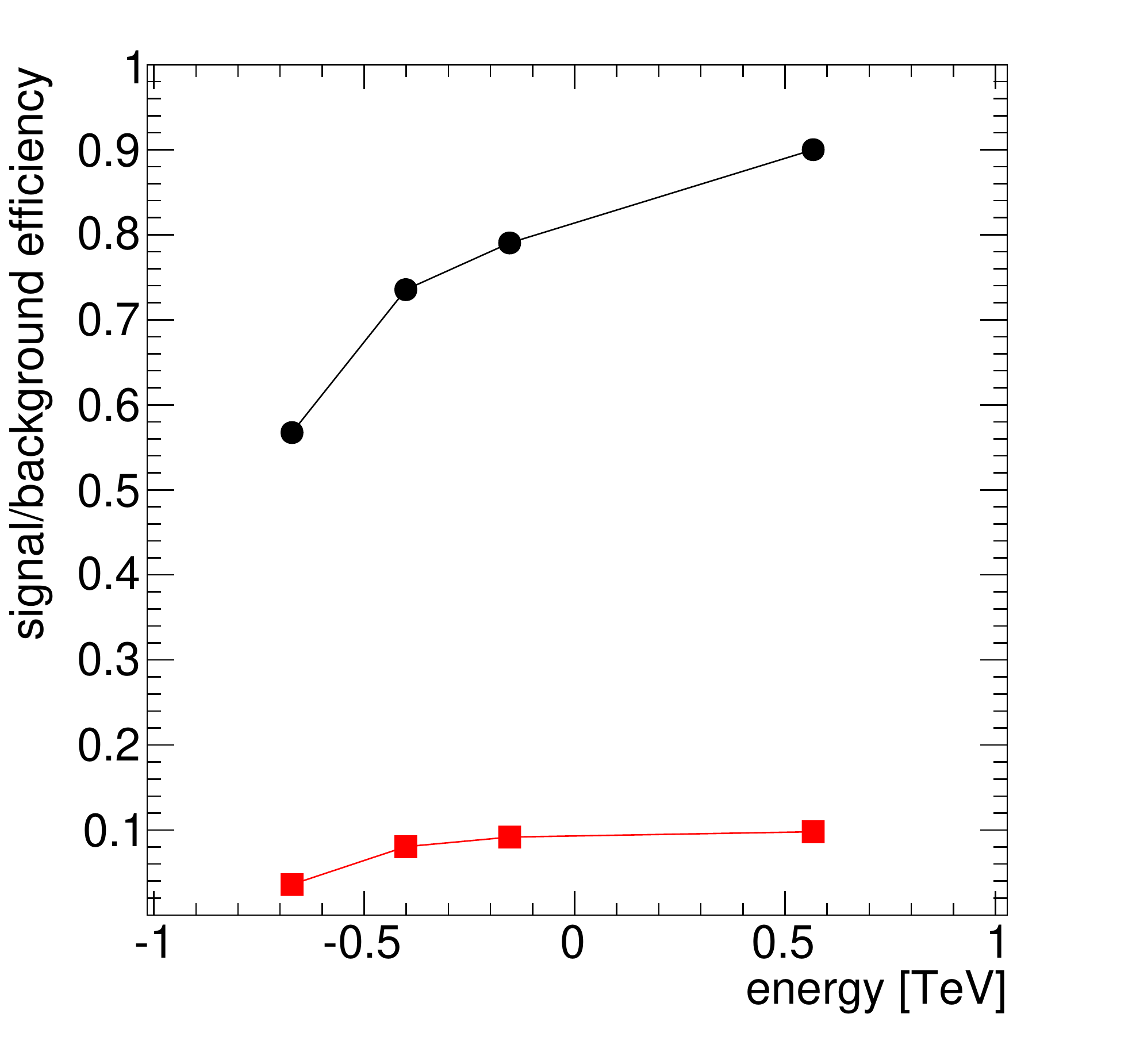}}
\caption{The signal (black curve, circles) and background (red curve, squares) efficiencies as a function of energy after applying the optimal selection $\Tau_{sel}$, for zenith angles of 0$^{\circ}$--22.5$^{\circ}$. Where no error bars are visible, the uncertainties are smaller than the size of the symbol.}\label{fig:efficiencyvsenergy}
\end{figure}

The value of $\Tau_{sel}$ in each energy and zenith angle bin is determined assuming a source with the minimum strength necessary to be detected at the $S$=5$\sigma$ confidence level (where the significance $S$ and standard deviation $\sigma$ are calculated using the Li \& Ma likelihood ratio method; Eq. 17 in \cite{LiMa}) with at least 10 signal events. Signal and background selection efficiencies as a function of $\Tau_{sel}$ are scaled by realistic signal and background rates extracted from observations of the Crab Nebula after applying the preselection requirements described in Section~\ref{sec:sample}. The rates are multiplied by an assumed observation time of 20 h to optimize for detection of a strong source or 50 h to optimize for detection of a weak source. The resulting curves give the number of signal and background events in each energy and zenith angle bin as a function of $\Tau_{sel}$. The detection significance is calculated, and the value of $\Tau_{sel}$ that produces the maximum significance selected. Fig.~\ref{fig:efficiencyvsenergy} shows the signal and background efficiencies as a function of energy after applying the optimal selection $\Tau_{sel}$, for zenith angles of 0$^{\circ}$--22.5$^{\circ}$. The black (red) curve shows the signal (background) efficiency. As energy increases, it is possible to retain much of the signal ($>$70\%) while suppressing the majority of the background (90\%).

Following the optimization of $\Tau_{sel}$, the cut on the parameter $Size2Max$ was set to ensure an energy threshold (defined as the energy at which the average energy bias falls below 10\%) similar to that of the standard VERITAS analysis. For the current VERITAS instrument configuration, the energy threshold for observations taken at a zenith angle of 20$^{\circ}$ is $\sim$170 GeV for soft cuts, $\sim$205 GeV for moderate cuts, and $\sim$350 GeV for hard cuts, after the application of the cuts on $\Tau_{sel}$ and $Size2Max$. These values can be compared with the energy threshold for the standard analysis: $\sim$165 GeV for soft cuts, $\sim$200 GeV for moderate cuts, and $\sim$345 GeV for hard cuts. The differences in the energy threshold for the two analyses are not significant and consequently do not significantly impact the performance studies shown below.
 
\section{Comparison between data and simulations}

The response of the multivariate analysis is compared between simulations and data excess events. The dataset used for this study is a subset of Crab Nebula observations. It contains a total livetime of \unit{14.2}{h}. The offset of the observations is \unit{0.5}{\degree} from the camera center which is also the case for the simulations. The zenith angles range from \unit{15}{\degree} to \unit{25}{\degree}. They are compared to $\gamma$-ray simulations at \unit{20}{\degree}.

\begin{figure}[t]
\centering
\subfigure{\includegraphics[width=\textwidth]{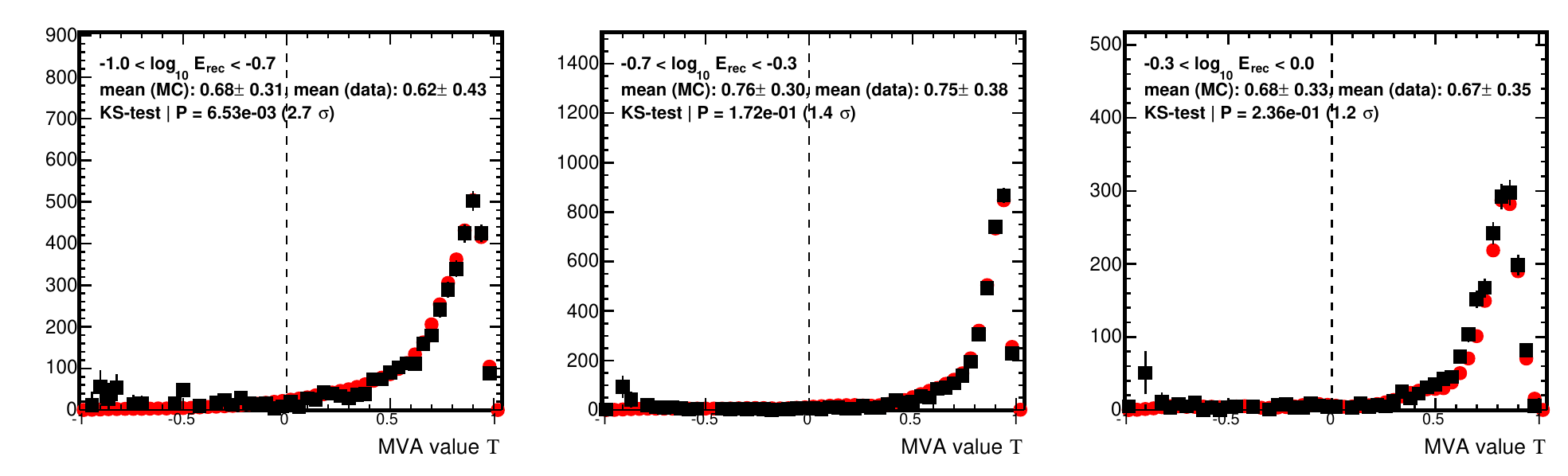}} \\
\subfigure{\includegraphics[width=\textwidth]{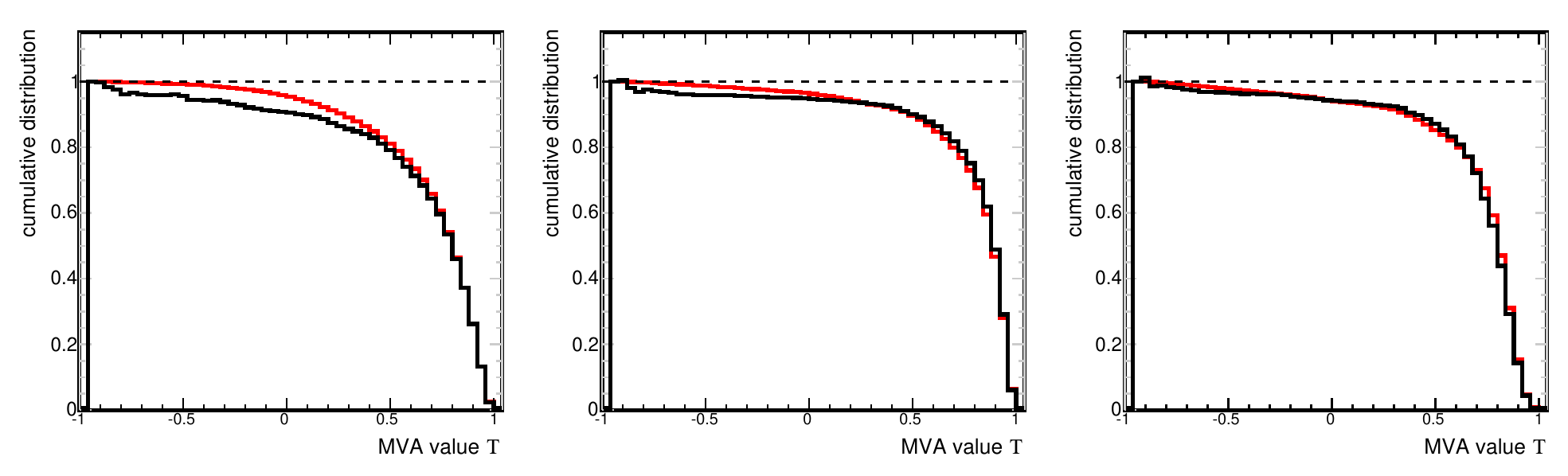}}
\caption{Comparison of the MVA distribution for $\gamma$-ray simulations (red points) and $\gamma$-ray excess events (black squares) dependent on energy (upper three plots). The corresponding cumulative distributions (three lower plots) of $\gamma$-ray simulations (red line) and data (black line) are also shown.}
\label{MVA energy dependent dist}
\end{figure}

Fig. \ref{MVA energy dependent dist} represents an energy-dependent comparison between $\gamma$-ray excess events and simulated $\gamma$-rays. The agreement between data and simulations demonstrates that the BDT classifies both simulated and real events in a similar way over the studied energy range. The cumulative distributions (three lower plots of Fig. \ref{MVA energy dependent dist}) show a disagreement of about 5\% between data and Monte-Carlo simulations which is small in comparison to other systematic uncertainties. 

There is a small bump in the data at $\Tau=-0.9$, most probably due to poorly reconstructed events. This effect will be the subject of future work.

\section{Performance of the boosted decision tree analysis}\label{sec:performance}
In the VERITAS standard analysis (box cuts) \cite{Daniel2007, Cogan2007}, suppression of background events is accomplished by placing selection requirements individually on the shower and image properties, namely $MRSW$, $MRSL$, $Size2Max$, $\chi^{2}$(E), and the emission height. As described above, these properties are included in the training parameters used for the BDT analysis. For both box and BDT analysis, a further cut is placed on $\theta^{2}$, the squared angular distance between the reconstructed arrival direction of the showers and the estimated location of the object. The box analysis cuts are optimized in a similar manner to the BDT cuts. A scan over each of the selection parameters is performed for a range of source strengths and values are selected that maximize the significance.

\subsection{\texorpdfstring{Quality factor for BDT selection}{Quality factor for BDT selection}}

A parameter commonly used in astronomy to quantify the performance of analysis cuts is the quality factor $q$ \cite{Bugayov2001}, defined as

\begin{equation}
q=\frac{\epsilon_{\gamma}}{\sqrt{\epsilon_{CR}}}
\label{eq: qfactor}
\end{equation}
with $\epsilon_{i}=\hat{N}_{i}/N_{i}$ and $i$ denoting $\gamma$- or cosmic ($CR$) rays. $\hat{N}_{i}$ and $N_{i}$ are the number of events after and before applying the selection criteria. The average $q$-factor ratio $q_{\Tau}/q_{box}$ is 1.17 for soft cuts. Calculating the $q$-factor ratio in zenith angle and energy bins indicates consistent performance across all bins.

\subsection{\texorpdfstring{Sensitivity of BDT analysis}{Sensitivity of BDT analysis}}
\begin{figure}[H]
\centerline{\includegraphics[width=0.5\textwidth]{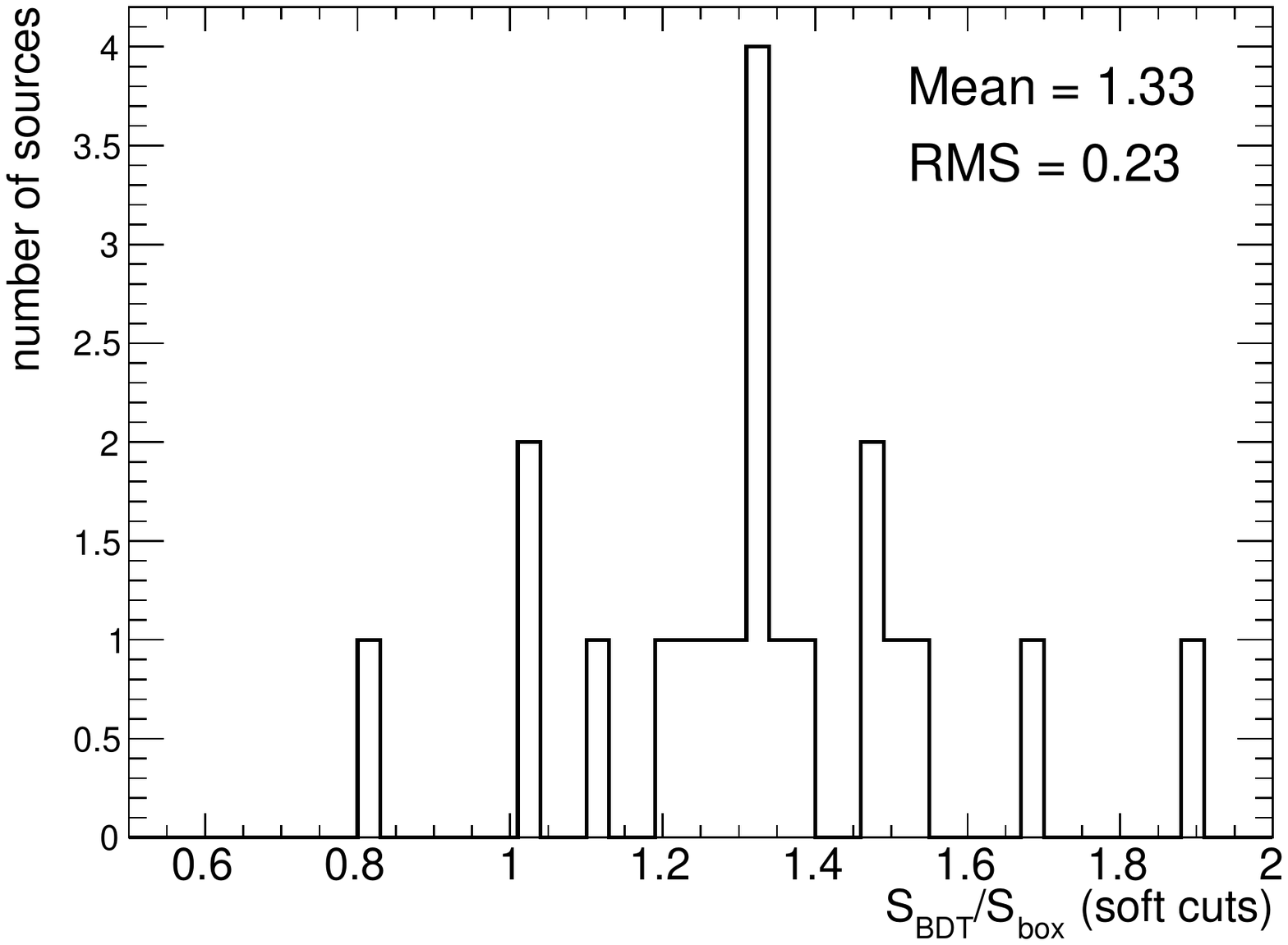}
\includegraphics[width=0.5\textwidth]{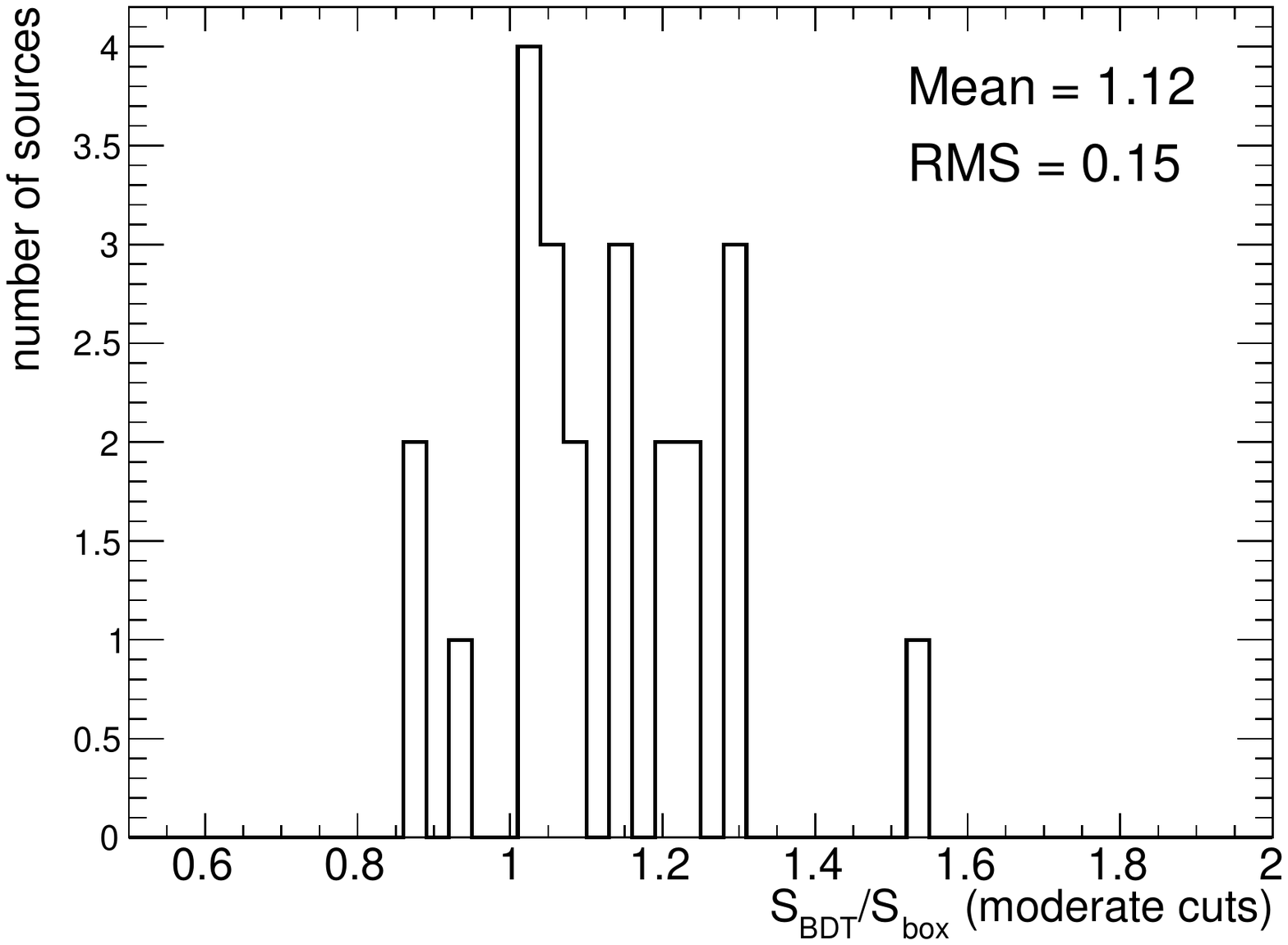}}
\caption{Comparison of detection significance for soft (left panel) and moderate (right panel) box and BDT cuts for a set of known VHE emitters. Objects detected above 3$\sigma$ with standard box cuts were considered, resulting in 20 and 23 sources for soft and moderate cuts, respectively.}\label{fig:NextDaySoft}
\end{figure}

The standard analysis of VERITAS consists of two independent packages, \textit{VEGAS} \cite{Cogan2007} and \textit{eventdisplay} \cite{Daniel2007}. Here, results obtained with the \textit{eventdisplay} package are studied. Furthermore, the results shown here focus on the array configuration after 2012. The performance of the BDT selection was compared to the performance of the standard box cut selection for a number of sources with different spectral properties, and demonstrates an average improvement in the significance of the detection of the object. This is shown in Fig.~\ref{fig:NextDaySoft}. The detection significance $S$ of known VHE sources was compared for soft and moderate box and BDT cuts. Datasets where the objects were detected above 3$\sigma$ with standard box cuts were used, resulting in 20 sources for soft cuts and 23 sources for moderate cuts. For soft cuts, out of the 20 objects tested, 19 showed equal or better performance with the BDT analysis (defined as $S_{BDT}/S_{box} \geq 1$). For moderate cuts, 20 out of 23 sources showed $S_{BDT}/S_{box} \geq 1$. It appears based on the mean of the distributions that the average improvement from the BDT analysis is greater with soft cuts than with moderate cuts, but the small sample size does not allow a firm conclusion. 

\begin{figure}[t]
\centerline{
\includegraphics[width=0.49\textwidth]{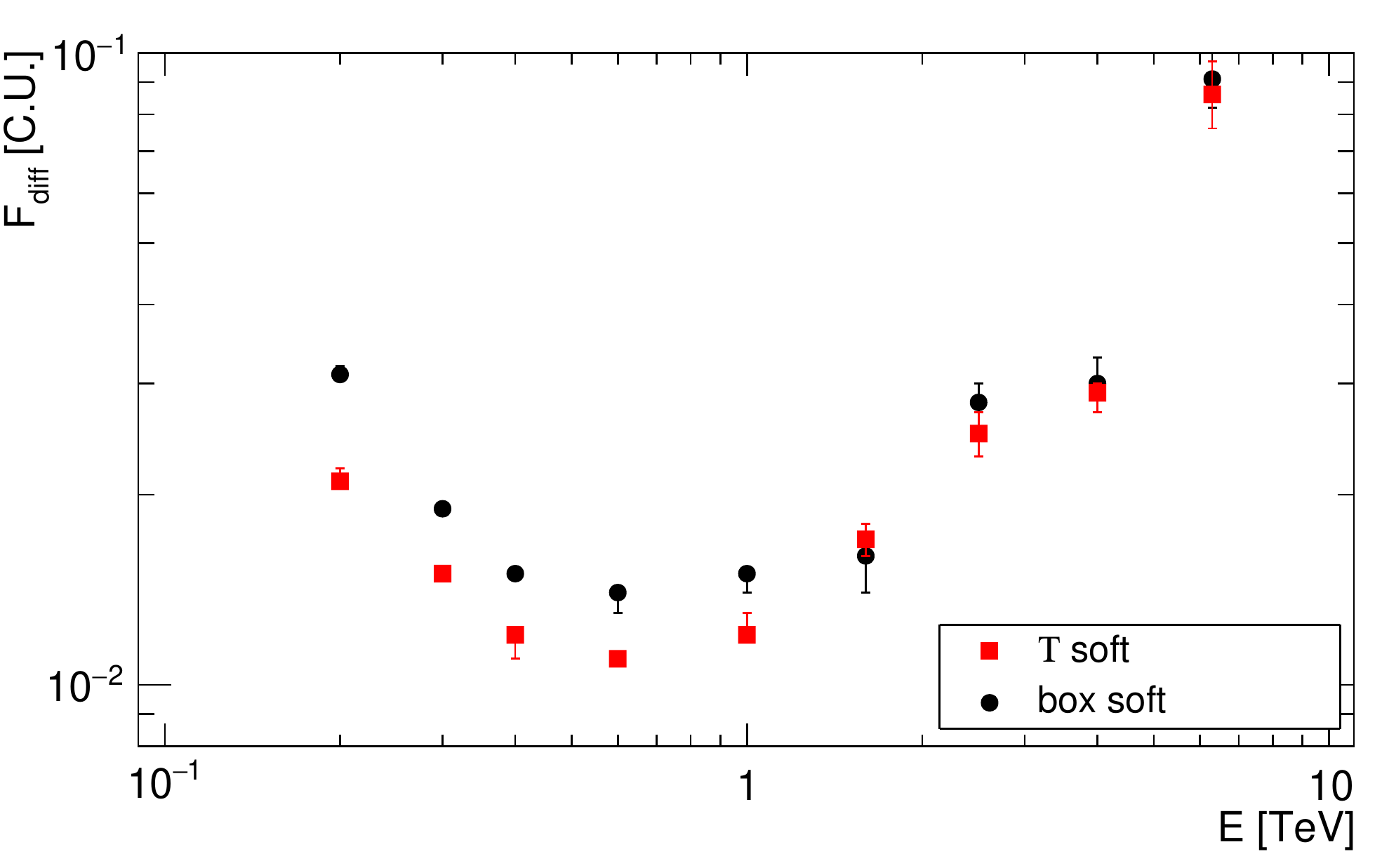}
\includegraphics[width=0.49\textwidth]{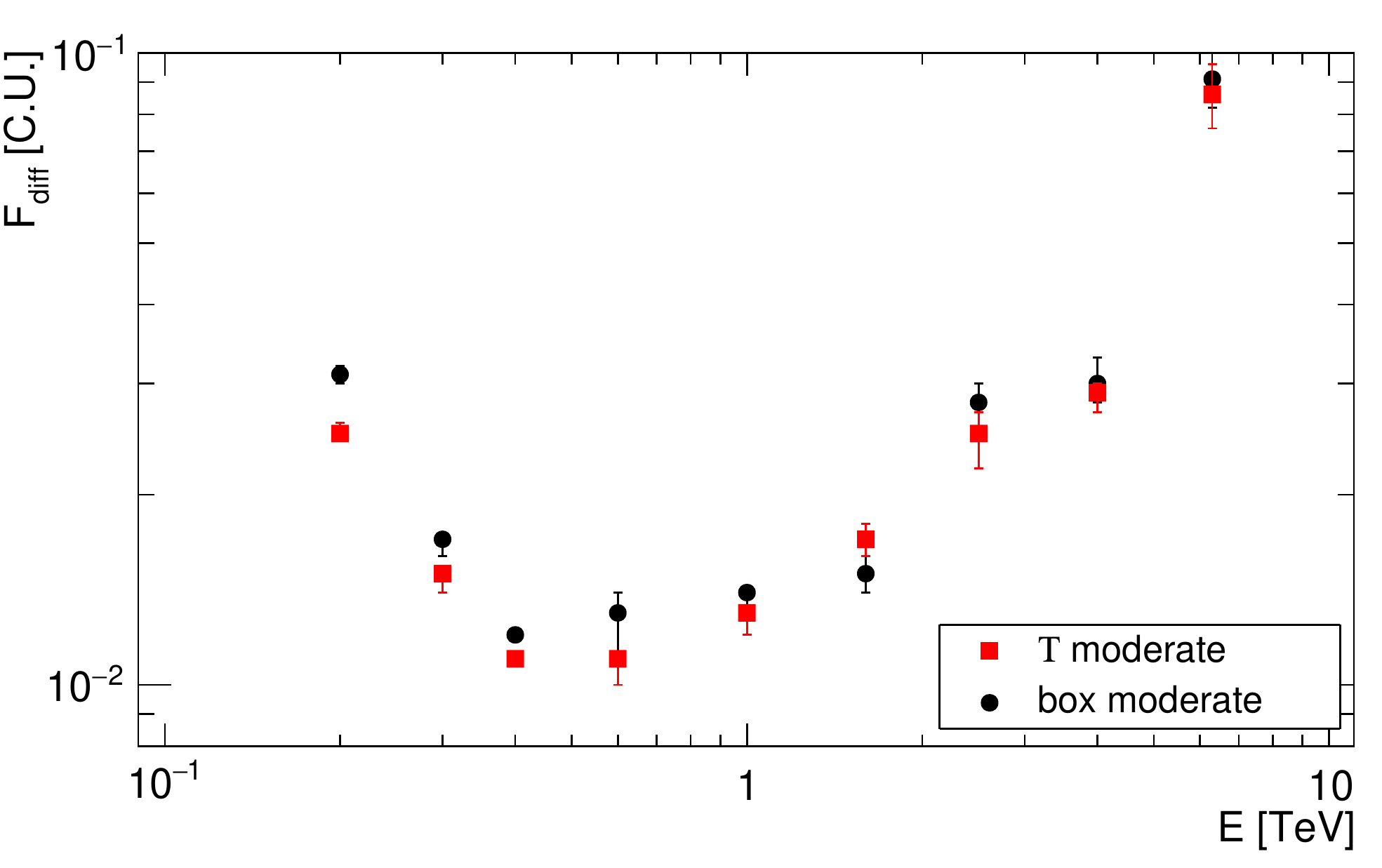}}
\caption{Comparison of differential flux sensitivity, $F_{diff}$, at a given energy, $E$, between BDT versus standard box selection using soft (left panel) and moderate (right panel) cuts, for high elevation Crab Nebula data, assuming 50 h of observing time. The data are binned in five energy bins per decade, evenly spaced in the logarithm of the energy. At least 5$\sigma$ detection significance and 10 excess events are required per bin. The differential flux is given in Crab units (C.U.).}
\label{fig:sensitivitycomparison}
\end{figure}

A benchmark test of the BDT performance compared to the standard box analysis~\cite{Park:2015vp} was performed by analyzing Crab Nebula data with soft and moderate cuts in terms of differential flux sensitivity. The differential sensitivity represents the lowest flux in a given energy bin which results in a significant detection after 50 h of observation. It is calculated in five energy bins per decade. A signal-free background region six times larger than the signal region ($\alpha=1/6$) are assumed in the following. The basic requirements for a significant detection per energy bin are a statistical significance of 5$\sigma$ and at least 10 excess events. Crab Nebula data taken at high elevation (zenith angles <  20$\degree$) were used in order to allow comparison of the two methods at the lowest energies. The results are shown in Fig.~\ref{fig:sensitivitycomparison}, Table~\ref{tab:RatesSensitivityPlotsoft}, and Table~\ref{tab:RatesSensitivityPlotmoderate}. Sensitivity improvements can be seen across the energy range, with the most significant improvements at lower energies. 

\begin{table}[!htb]
\centerline{
\begin{tabular}{|c|c||c|c||c|c|}
\hline
$E_{min}$    &  $E_{max}$   &   $N_{ON}$ Box   &   $N_{ON}$ BDT   &   $N_{OFF}$ Box   &   $N_{OFF}$ BDT\\
\hline
\hline
\unit{0.13}{TeV} 	&   \unit{0.20}{TeV}  	&   10553$\pm$103   &   6816$\pm$83   &   16429$\pm$128   	&   4544$\pm$67 \\ 
\unit{0.20}{TeV}   &   \unit{0.32}{TeV}     &   7411$\pm$861   	&   6506$\pm$81   &   4240$\pm$65   		&   2189$\pm$47 \\
\unit{0.32}{TeV}   &   \unit{0.50}{TeV}     &   5338$\pm$73   		&   5196$\pm$72   &   1456$\pm$38   		&   840$\pm$29 \\
\unit{0.50}{TeV}   &   \unit{0.80}{TeV}   	&   3443$\pm$59   		&   3586$\pm$60   &   499$\pm$22   		&   344$\pm$19 \\
\unit{0.80}{TeV}   &   \unit{1.26}{TeV}    	&   1885$\pm$43   		&   1996$\pm$45   &   155$\pm$12   		&   114$\pm$11 \\
\unit{1.26}{TeV}   &   \unit{2.00}{TeV}    	&   984$\pm$31   		&   947$\pm$31   	&   36$\pm$6   			&   40$\pm$6 \\
\unit{2.00}{TeV}   &   \unit{3.16}{TeV}    	&   459$\pm$21   		&   468$\pm$22  	&   21$\pm$5   			&   17$\pm$4 \\
\unit{3.16}{TeV}   &   \unit{5.00}{TeV}   	&   239$\pm$16   		&   254$\pm$16   	&   1$\pm$1   				&   1$\pm$1 \\
\unit{5.00}{TeV}   &   \unit{7.94}{TeV}    	&   80$\pm$9   			&   89$\pm$9   		&   1$\pm$1   				&   3$\pm$2 \\
\hline
\end{tabular}
}
\caption{Comparison of the number of events for the ON- ($N_{ON}$) and OFF-region ($N_{OFF}$) for each energy bin of Fig.~\ref{fig:sensitivitycomparison} (left panel), derived from about 36 h of high elevation observations of the Crab Nebula.}
\label{tab:RatesSensitivityPlotsoft}
\end{table}

\begin{table}[!htb]
\centerline{
\begin{tabular}{|c|c||c|c||c|c|}
\hline
$E_{min}$    &  $E_{max}$   &   $N_{ON}$ Box   &   $N_{ON}$ BDT   &   $N_{OFF}$ Box   &   $N_{OFF}$ BDT\\
\hline
\hline
\unit{0.13}{TeV} 	&   \unit{0.20}{TeV}  	&   4324$\pm$66   		&   3490$\pm$59   &   3523$\pm$60   		&   1694$\pm$41 \\ 
\unit{0.20}{TeV}   &   \unit{0.32}{TeV}     &   6308$\pm$79   		&   5900$\pm$77   &   2489$\pm$50   		&   1699$\pm$41 \\
\unit{0.32}{TeV}   &   \unit{0.50}{TeV}     &   4958$\pm$70   		&   5042$\pm$71   &   835$\pm$29   		&   739$\pm$27 \\
\unit{0.50}{TeV}   &   \unit{0.80}{TeV}   	&   3370$\pm$58   		&   3565$\pm$60   &   337$\pm$18   		&   318$\pm$18 \\
\unit{0.80}{TeV}   &   \unit{1.26}{TeV}    	&   1873$\pm$43   		&   1995$\pm$45   &   131$\pm$11   		&   113$\pm$11 \\
\unit{1.26}{TeV}   &   \unit{2.00}{TeV}    	&   983$\pm$31   		&   947$\pm$31   	&   35$\pm$6   			&   40$\pm$6 \\
\unit{2.00}{TeV}   &   \unit{3.16}{TeV}    	&   459$\pm$21   		&   468$\pm$22  	&   21$\pm$5   			&   17$\pm$4 \\
\unit{3.16}{TeV}   &   \unit{5.00}{TeV}   	&   239$\pm$16   		&   254$\pm$16   	&   1$\pm$1   				&   1$\pm$1 \\
\unit{5.00}{TeV}   &   \unit{7.94}{TeV}    	&   80$\pm$9   			&   89$\pm$9   		&   1$\pm$1   				&   3$\pm$2 \\
\hline
\end{tabular}
}
\caption{Comparison of the number of events for the ON- ($N_{ON}$) and OFF-region ($N_{OFF}$) for each energy bin of Fig.~\ref{fig:sensitivitycomparison} (right panel), derived from about 36 h of high elevation observations of the Crab Nebula.}
\label{tab:RatesSensitivityPlotmoderate}
\end{table}

\subsection{\texorpdfstring{Spectral reconstruction with BDT analysis}{Spectral reconstruction with BDT analysis}}

\begin{figure}[t]
\centerline{\includegraphics[width=0.65\textwidth]{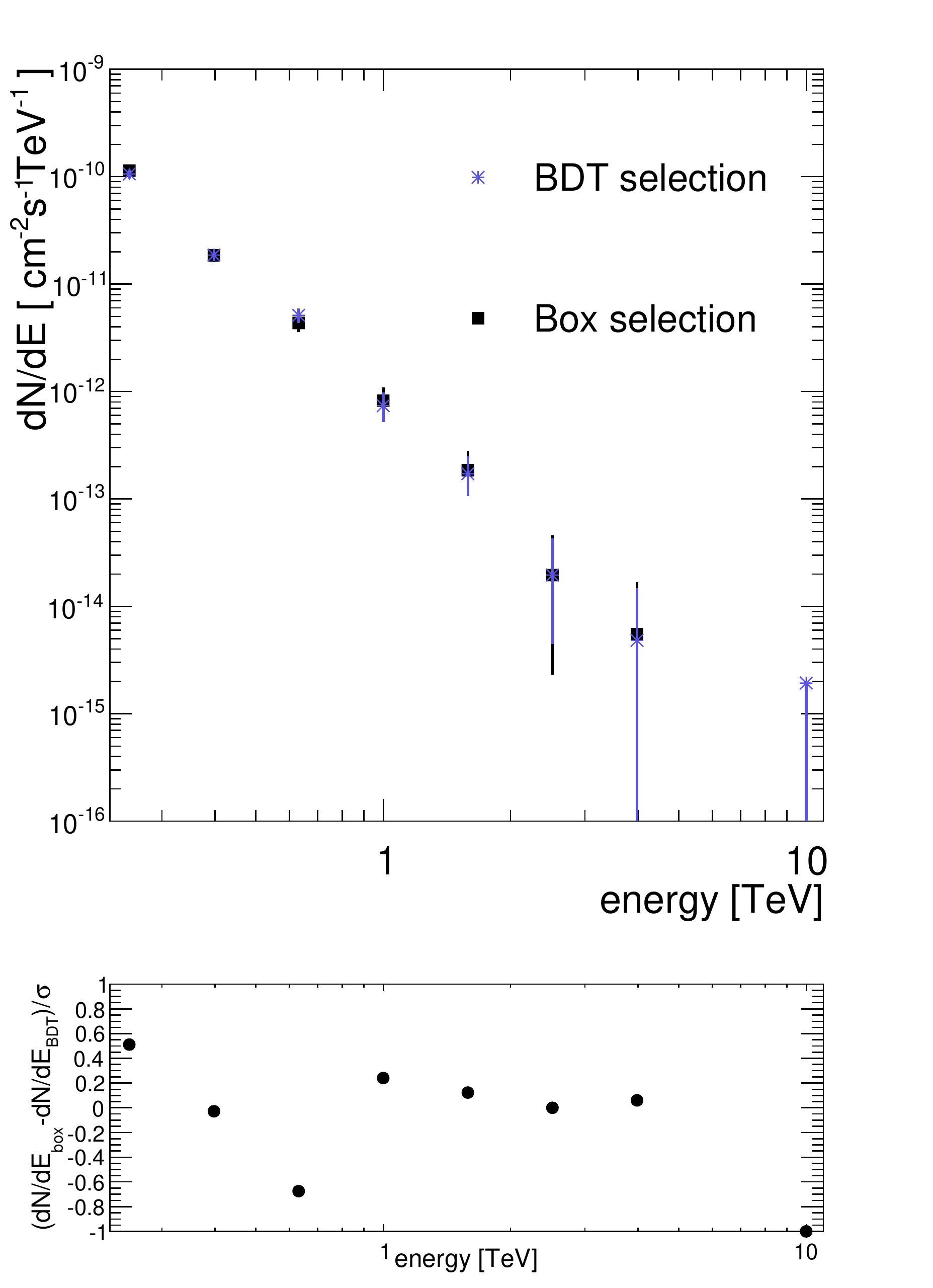}}
\caption{Comparison of energy spectra using BDT versus standard box selection cuts, for the blazar \textit{VER~J0521+211}. The lower plot shows the residuals.}\label{fig:spectralcomparison}
\end{figure}

Energy spectra were compared for BDT versus standard box selection, to verify that no spectral features were introduced by the BDT analysis. This was of particular concern, given that the selection on the BDT response is optimized in each energy and zenith angle bin individually, leading to 16 different $\Tau_{sel}$ values. Fig.~\ref{fig:spectralcomparison} shows the energy spectra for the blazar \textit{VER~J0521+211}~\cite{VERJ0521} for moderate BDT and standard box selection cuts. The differential flux points differ by less than one standard deviation between the methods for all points, as illustrated by the residuals in the lower panel.

\subsection{\texorpdfstring{Performance on non-standard datasets}{Performance on non-standard datasets}}
As discussed in Section~\ref{sec:sample}, the BDTs were trained with data and simulation that used a 0.5$^{\circ}$ wobble offset and had four telescopes operational. It was thus necessary to test the performance of the BDT analysis on datasets taken with large wobble offsets and with only three telescopes operational.

To verify that the training is accurate for datasets with wobble offsets not equal to 0.5$\degree$, BDT performance on Crab Nebula observations with wobble offsets of 0.7$\degree$, 1.0$\degree$, and 1.3$\degree$ was tested. The performance of the moderate BDT selection is compared against moderate standard box cuts. The results are shown in Table \ref{tab:wobble3Teltest}. An improvement in significance and background rate compared to standard box cuts is observed regardless of the wobble offset of the data. However, a dedicated training for data taken at large offsets could be the subject of future work.

\begin{table}[t]
\centerline{
\begin{tabular}{|c|c||c||c|c||c|c|}
\hline
offset   &   NTel   &  $S_{BDT}$/$S_{Box}$   &   $\gamma$/min Box   &   $\gamma$/min BDT   &   bkg/min Box   &   bkg/min BDT\\
\hline
\hline
0.5$\degree$   &   4   &   1.07   &   7.83$\pm$0.09   &   7.89$\pm$0.09   &   0.39$\pm$0.02   &   0.62$\pm$0.02 \\ 
0.7$\degree$   &   4   &   1.08   &   7.57$\pm$0.22   &   8.44$\pm$0.23   &   0.57$\pm$0.06   &   0.54$\pm$0.05 \\
1.0$\degree$   &   4   &   1.14   &   4.97$\pm$0.17   &   5.92$\pm$0.19   &   0.45$\pm$0.05   &   0.42$\pm$0.04 \\
1.3$\degree$   &   4   &   1.14   &   2.90$\pm$0.13   &   3.33$\pm$0.14   &   0.29$\pm$0.04   &   0.24$\pm$0.04 \\
0.5$\degree$   &   3   &   1.15   &   3.03$\pm$0.11   &   3.80$\pm$0.12   &   0.17$\pm$0.03   &   0.17$\pm$0.03 \\
\hline
\end{tabular}
}
\caption{Performance of training with 0.5$\degree$ wobble offset and four telescopes operating applied to Crab Nebula data taken with different wobble offsets/only three telescopes operating. BDT performance is compared against moderate box cuts.}
\label{tab:wobble3Teltest}
\end{table}

The last row of Table \ref{tab:wobble3Teltest} shows the performance of moderate BDT and moderate box cuts on Crab Nebula data taken with only three telescopes operational. While the $\gamma$-rate increases by about 25\% from moderate box to BDT cuts, the background rate is comparable. 

\section{Conclusions}
In this study, the training and evaluation of BDTs for selecting $\gamma$-ray events from VERITAS data was presented. This method combines the information carried in several parameters to classify $\gamma$-ray- or cosmic-ray-like images based on a single parameter $\Tau$. The value of this parameter, for a given event, is used to classify events as being of electromagnetic or hadronic origin. Energy- and zenith angle-dependent cuts are introduced to account for the dependency of the BDT training variables on the reconstructed shower energy and zenith angle of observations. Our results clearly show that a multivariate approach using BDTs increases the sensitivity of VERITAS for a large variety of sources. It is envisaged that future work will use an expanded background training sample that allows separate training for observations of southern and northern astrophysical objects, and for winter and summer observations.

\section*{Acknowledgements}
The authors would like to thank the VERITAS collaboration for the available data. The authors additionally acknowledge the support of their host institutions. M.K. and G.M. acknowledge support from the Helmholtz Alliance for Astroparticle Physics. E.P. acknowledges the funding provided by a Marie Curie Intra-European Fellowship. The authors thank M.K. Daniel, H. Fleischhack, K.J. Ragan, and D. Williams for helpful conversations.

\section*{References}

\bibliography{mybibfile}

\end{document}